# Efficient high-quality photon pair generation in modal phase-matched thin-film lithium niobate micro-ring resonators

Tingting Chen[1], Feihong Xue[1], Ryan Hogan[1], Xiaofei Ma[1], Jiaxuan Zhou[1], Yule Zhao[1], Yanling Xiao[1], Zhilin Ye[1,2], Chong Sheng[1], Qiang Wang[1], Shining Zhu[1], and Hui Liu[1,*]

[1]*National Key Laboratory of Solid State Microstructure, School of Physics, and Collaborative Innovation Center of Advanced Microstructures, Nanjing University, Nanjing, 210093, China*

[2]*Nanzhi Institute of Advanced Optoelectronic Integration, Nanjing 211800, China*

\* Correspondence authors: liuhui@nju.edu.cn

Homepage: https://einstein.nju.edu.cn/index_en

## Abstract

**Efficient generation of high-quality photon pairs is essential for modern quantum technologies. Micro-ring resonator is an ideal platform for studying on-chip photon sources due to strong nonlinear effect, resonant-enhanced optical fields, and high integration. Thin-film lithium niobate (TFLN) micro-ring resonators with periodically poled quasi-phase matching have shown high-quality photon pair generation. However, periodic poling technology remains expensive and requires complex fabrication hindering its scalability and capability for practical application in nonlinear photonic devices. To address this, we propose a scalable approach using TFLN micro-ring resonators based on modal phase matching to achieve cost-effective, efficient high-quality photon-pair generation, significantly simplifying fabrication. We achieved pair generation rates up to 40.2 MHz/mW through spontaneous parametric down-conversion, with coincidence-to-accidental ratios exceeding 1200. By combining micro-ring resonance enhancement with modal phase matching, our approach reduces device size and fabrication cost while maintaining high nonlinear efficiency. These results advance the development of compact, efficient on-chip photon sources for next-generation nonlinear and quantum photonic applications.**

# 1  Introduction

Modern quantum information technologies exploit the utility of single photons for quantum computation, quantum communication, and metrology, demonstrating superior advantages over classical approaches [1]-[3]. Quantum-encoded encryption keys, for example, can fundamentally enhance communication security, utilizing quantum photon properties to achieve unprecedented measurement precision [4]. Quantum protocols to realize these processes rely on an efficient and stable photon source as a vital component, and as a result, extensive research has been conducted on single photon sources, including spontaneous parametric down-conversion (SPDC) in bulk crystals, optical waveguides, metasurfaces, and two-dimensional materials [5]-[9], spontaneous four-wave mixing (SFWM) in optical fibers and silicon materials [10], [11], and quantum dot exciton processes in semiconductor materials [12]. Quantum dots have been shown to generate single photons or photon pairs through exciton cascade recombination, enabling deterministic single-photon generation, but require low-temperature and resonant pumping [13]. Both SPDC and SFWM, however, use nonlinear optical processes to generate correlated photon pairs, offering inherent advantages. Particularly, as discussed in [14], these inherent advantages include room-temperature operation, simple signal filtering, coherent emission characteristics, multi-degree-of-freedom entanglement capabilities, and inherent compatibility with photonic integration platforms. Notably, with a higher nonlinear coefficient, SPDC process requires lower pump light power compared to the SFWM process, making it a primary method for generating photon pairs.

LN is a multifunctional crystal with excellent optical properties such as wide transparent window, low optical loss, large second-order nonlinear coefficient, and long-term stability, making it an ideal material system for integrated optics and nonlinear optics [15]. Various nonlinear and quantum photonic devices have been realized based on LN material, such as efficient frequency conversion using second-harmonic generation (SHG), sum- frequency generation (SFG), and difference-frequency generation (DFG) [16]-[20], supercontinuum generation [21], [22], optical frequency combs [23]-[25], etc. Nevertheless, second-order nonlinear effects in LN and subsequent periodic poling can serve as a platform for efficient photon sources [26]. LN processing is still undergoing optimization with recent developments in micro-fabrication technology. In particular, the tightly confined thin film lithium niobate (TFLN) ridge waveguide platform shows significant progress with respect to its early counterparts regarding weakly confined proton exchange/titanium diffusion waveguides. These technological advances not only improve the yield of photon pairs but also greatly promote quantum optical integrated chip development [27]-[30].

As one of the important devices in the field of inte-grated optics, micro-ring resonators (MRRs) have wide applications in filtering [31], modulation [32], all-optical switches [33], lasers [34], and other fields. Compared to straight waveguides, MRRs are advantageous due to their small size and resonant enhancement, which can induce strong nonlinear optical processes, thereby reducing the requirements for a long waveguide length and larger pump intensity. TFLN MRRs made of TFLN materials based on periodic poling have shown efficient SPDC and generation of high-quality photon pairs [35]-[38]. However, the preparation technology of quasi-phased matching based on periodic poling is complex and costly, which hinders the large-scale development and application of nonlinear devices. In contrast, modal phase matching does not comprise complex fabrication processes and is easy to integrate on-chip. Moreover, resonant enhancement from MRRs reduces the requirement for the system's intrinsic nonlinear strength and circumvents the need for complex and costly periodic poling technology. The research on modal phase-matching in TFLN MRRs, however, remains relatively unexplored, especially in the context of photon-pair generation.

In this work, we design and fabricate TFLN MRRs based on modal phase matching to achieve efficient high-quality photon-pair source on chip. We also design and implement an asymmetric coupler as a mode converter to help meet phase-matching conditions for the pump light mode. Our fabricated chips achieved Q-factors exceeding $10^5$ and met triple-resonance conditions through precision temperature stabilization and wavelength selection. As a result, we experimentally observed an overall pair generation rate (PGR) as high as 40.2 MHz/mW, with a coincidence-to-accidental ratio (CAR) higher than 1200. Overall, the TFLN MRR modal-phase matched platform shows promise for an overall smaller footprint, easier photonic integration, and simplified fabrication processes, proving beneficial to future development advancements in nonlinear and quantum optical technologies.

## 2 Structural design and numerical calculation

A schematic of the designed TFLN MRR structure on chip is shown in Figure 1(a), consisting of two main parts: the MRR and the mode converter. The mode converter converts 775 nm light from the TM00 mode to the TM20 mode. The mode converter is then paired with the corresponding MRR that satisfies phase matching conditions of two light waves near 775 nm and 1550 nm. As a result, efficient high-quality photon pair generation can be achieved through SPDC in the telecommunications frequency band, as shown in Figure 1(b).

### 2.1 Photon pair generations based on modal phase matching

We employ Z-cut TFLN to ensure efficient nonlinear conversion, where light propagates along the circumference of the MRR and the crystal's optical axis is oriented perpendicular to the light propagation direction. This alignment ensures the preservation of Type I collinear phase matching. In contrast, using X-cut TFLN—with a horizontal optical axis—introduces a continuously changing angle between the propagation direction and the optical axis, thereby hindering efficient nonlinear processes [39].

To achieve efficient nonlinear processes in the MRR, both phase matching and resonance conditions must be simultaneously met. The required phase matching condition must satisfy $\Delta k = k_p - k_s - k_i = 0$, where $k_p$, $k_s$, and $k_i$ are the propagation constants of the pump, signal, and idler light, respectively. Here, we choose the TM20 mode for the pump light and TE00 mode for the signal and idler light to achieve modal phase matching. Simultaneously, we require that the MRR free spectral range (FSR) is not too large to ensure all interacting beams can resonate. Therefore, we choose a minimal micro-ring radius of 150 μm, corresponding to a FSR of 1.1 nm. Under these conditions, along with temperature control and previse wavelength selection, the pump, signal, and idler can simultaneously achieve resonance with ease. To ensure properly control, we calculate the waveguide mode dispersion of the MRR.

Figure 1(c) shows the MRR waveguide cross-section. We choose a TFLN layer thickness of 300 nm, etching depth of 200 nm, and the waveguide sidewall inclination angle of 70°. Here, we consider the change in the effective refractive index caused by waveguide bending. The variation of the effective refractive index of TM20 mode at 775 nm and TE00 mode at 1550 nm are shown in Figure 2(a). Here, we vary the top width of the 150-μm-radius MRR waveguide. From Figure 2(a), it can be seen that the TM20 mode at 775 nm and the TE00 mode at 1550 nm satisfy modal phase matching when the top width is 1.21 μm. Figure 2(b) shows that for a top width of 1.21 μm, the pump wavelength satisfying phase matching also increases with increase of the MRR radius. However, only a 0.1 nm change in the required pump wavelength for phase matching is observed when varying the radius from 150 μm to 400 μm.

## 2.2 Mode converter design

We designed and integrated a mode converter at the front end of the micro-ring structure to achieve modal phase matching in the MRR system, such that the pump light operates in the TM20 mode and the input straight waveguide supports the fundamental mode,TM00 (See Figure 1(a)). As shown in Figure 2(c), the effective refractive indices of the TM00 and TM20 modes of 775 nm pump light were calculated under different waveguide widths. Effective energy exchange can be achieved between these two modes when their effective refractive indices in each waveguide are equal. As a result, we chose top widths of 0.84 μm and 3 μm, for the mode converter input and output waveguide, respectively, as depicted by the red lines of Figure 2(c). The coupling gap between these waveguides was set to be 0.2 μm, making an overall compact structure. To achieve complete mode conversion, a coupling length of 600 μm is required. Using a coupling length of 600 μm, we calculated the coupling efficiency of different pump wavelengths using FDTD, plotted in Figure 2(d), resulting in a maximum coupling efficiency of 98% within the range of 700 – 850 nm. The field distribution at optimal energy transfer is shown in the insert in Figure 2(d).

## 3 Experimental results

## 3.1 Sample preparation and characterization

The fabricated sample consists of commercial Z-cut silicon-based TFLN (NANOLN Inc.), with a 525-μm-thick silicon bottom-layer, a 2.025-μm-thick silicon dioxide middle layer, and a 300-nm-thick TFLN layer on top. The sample fabrication process primarily involves electron beam lithography and ion beam etching, thereby avoiding the complexity associated with periodically poled waveguides. From here, MRRs with different coupling gaps and radii were prepared. One sample set varied the coupling gap from 0.3 μm to 0.5 μm in 0.05 μm increments while holding a constant radius of 150 μm. The other sample set held a constant coupling gap of 0.3 μm while varying the radius from 150 μm to 300 μm with 50 μm increments. After etching, a polish of 150 μm was applied to both ends of the chip to reduce end face roughness. Figure 3 shows the detailed analysis of the fabricated samples.

The MRR microscopy image is shown in Figure 3(a), and SEM images of the coupler and waveguide are shown in Figure 3(b) and (c). The measured transmission spectrum of the sample without a mode converter with 150 μm radius and 0.45 μm coupling gap is shown in Figure 3(d). Under these conditions, we observed critical coupling and measured the FSR to be 1.1 nm, consistent with our theoretical calculation. We found that the propagation loss is approximately 1.36 dB/cm at 1550 nm. The intrinsic quality factor can reach $2.97 \times 10^5$ and the measured loaded quality factor to be $1.28 \times 10^5$, as shown in Figure 3(e).

## 3.2 Second-harmonic generation

The experimental setup for SHG is shown in Figure 4(a). A continuously tunable laser (Santec TSL-570) is amplified by an erbium-doped fiber amplifier, directed through a polarization controller, and coupled into a waveguide using a fiber with a diameter of 3.2 μm. The output power and spectrum of the SHG signal is captured by a power meter and spectrometer. Note that the fundamental light is coupled directly, while the SHG signal is coupled with the mode converter for output, as depicted in Figure 4(a).

Figure 4(b) shows the transmission spectrum of the fundamental light from 1560 – 1580 nm for the critically coupled MRR, with a radius and coupling gap of 150 μm and 0.45 μm, respectively. The discrepancy between the transmission spectra presented here and those measured previously arises from the mode converter. The fundamental wavelength that satisfies the resonance condition of the MRR corresponds to the valley position on the transmission spectrum. Figure 4(c) shows the SHG signal measured in the range of 1560 – 1580 nm. It can be seen that in Figure 4(b) and (c), the SHG signal remains weak regardless of the resonant fundamental signal, as depicted by dashed lines outside the shaded regions. Only when fundamental signal and SHG signal are resonant and phase-matched, we achieve efficient SHG, as depicted by the shaded area in Figure 4(b) and (c). Careful tuning of the temperature allows for precise control of the optimal conditions. Figure 4(d) shows the second harmonic spectrum, with a maximum signal at $\lambda_{SHG} = 785.525$ nm, corresponding to a fundamental wavelength of $\lambda_{FF} = 1571.05$ nm. The

SHG and fundamental signals are measured using CCD cameras, as shown by the two inserts in Figure 4(d), indicating that both signals are resonant. Moreover, the CCD measurement shows that SHG is indeed generated in the MRR, rather than the straight waveguide. Finally, the power dependence of the SHG signal is measured by varying the input power of the fundamental plotted in Figure 4(e), following the expected quadratic dependence of the input fundamental power.

### 3.3 Photon pair generation

The schematic diagram of the experimental setup for photon pair generation and detection is shown in Figure 5(a). A near-visible wavelength tunable laser (SL-N-780-L-CW semiconductor laser) is passed through a polarization controller and then coupled through a 775 nm lens fiber into the tapered waveguide. The light is converted to a TM20 mode by the mode converter and then coupled into a MRR through the straight waveguide. Photon pairs are then generated from the MRR and coupled through a tapered waveguide to an output fiber with a diameter of 3.2 μm. After filtering the input pump light, the generated photon pairs are sent down two paths via a 50:50 beam splitter. The polarization of the light in each path is first controlled, and then subsequently detected on a superconducting nanowire single photon detector (SNSPD, detection efficiency~60%). The coincidence counts between the photon pairs are measured using a time-dependent single photon counter (TCSPC) (PicoQuant PicoHarp 300). The chip-to-fiber coupling losses are 10.7 dB around 780 nm and 6.2 dB around 1570 nm. The losses in each path from the output fiber to the detector are 7.47 dB and 6.68 dB at 1570 nm, respectively.

The photon pair generation was analysed for various MRRs with different coupling gaps and radii. Firstly, the coupling gap directly affects the pump light coupling efficiency into the MRR and the output signal coupling efficiency, thereby affecting the overall nonlinear conversion efficiency. As previously stated, optimal SHG efficiency occurs at a radius of 150 μm with a coupling gap of 0.45 μm. Therefore, one can expect similar peak in efficiency for photon pair generation under these conditions. Consequently, the average measured pump light wavelength for near-degenerate SPDC occurs at 785.21 nm, as shown in Figure 5(b). Note that, the change of gap width does not affect the phase matching condition, but as the coupling gap gradually increases, the coincidence count increases up to a maximum at a gap of 0.45 μm, and hereby after decreases, as shown in Figure 5(c). Figure 5(d) shows monotonic increase of the measured pump light wavelength for near-degenerate SPDC with respect to the MRR radius. The experimentally measured results are consistent with theory. Slight only differences can be attributed to minor fabrication errors affecting the resonance characteristics of the MRR.

Figure 6 shows a more detailed analysis of the photon pair generation. The spectral dependence of the pump light on pair generation is plotted in Figure 6(a) for a MRR with a 150 μm radius and a 0.45 μm coupling gap. Maximum coincidence counts were found to be at a pump wavelength of 785.44 nm. Note that the pump wavelength of the SPDC process only differs by approximately 0.1 nm from that of the SHG process. This discrepancy arises primarily from the inherent measurement uncertainty associated with the wavelength meter employed in the SPDC experiment. These results reveal a wavelength consistency between the two processes.

Figure 6(b) details about the measured rates and overall statistics of the source. Single photon and coincidence counts are measured as a function of input pump power and used to determine the on-chip photon pair generation rate (PGR) per unit of time, defined as:

$$\text{PGR} = \frac{N_1 N_2}{2 N_{12}} \tag{1}$$

where $N_{1(2)}$ is the single-photon detection count rate in each path and $N_{12}$ is the two-photon coincidence count rate. The brightness of a photon source quantified by its PGR, serves as a metric for evaluating performance as a quantum light source. Hence, the PGR at different pump powers is plotted in red, with the corresponding fit shown in black in Figure 6(b). When the pump light power is 12.3 µW, the coincidence count rate is found to be 167 Hz, and singles are found to exceed 14 KHz, the dark count is 1.5 KHz, resulting in a PGR up to 0.49 MHz. The slope of the fitting indicates the two-photon source brightness of 40.2 MHz/mW.

The coincidence-to-accidental ratio (CAR) plotted in blue in Figure 6(b) provides relevant information about the quality, and is defined as:

$$\text{CAR} = \frac{N_{CC} - N_{AC}}{N_{AC}} \tag{2}$$

where $N_{CC}$ is the number of total coincidence counts and $N_{AC}$ is the number of accidental coincidence counts. The maximum CAR was found to be 1786 at an input pump power of 6.14 µW. The CAR shows monotonic decrease with input pump power but remains above 1200 over a relatively wider range of input powers up to 12.3 µW. These results demonstrate the high quality of the generated photons from our source. Furthermore, we present a comparative assessment of MRRs fabricated from different materials for photon-pair generation using modal phase matching and continuous-wave pumping as shown in Table 1. Table 1 clearly indicates that the TFLN MRR developed in this work achieves exceptional brightness and CAR metrics, establishing their advantages as high-performance photon pair sources. Future improvements targeting the mode converter efficiency and quality of the chip facet coupling are expected to further elevate photon pair spectral brightness within the TFLN MRR.

## 4  Conclusions

We have demonstrated an on-chip integrated photon-pair source using a TFLN MRR based on modal phase matching. The MRR's structural parameters were engineered to achieve modal phase matching between telecommunication and near-visible wavelengths. Additionally, through optimization of the mode converter, the pump laser was configured to operate in higher-order mode under phase-matching condition during SPDC. The PGR was measured to be as high as 40.2 MHz/mW, with a CAR exceeding 1200 at input pump powers of 12.3 µW. These results confirm the efficiency and high-quality photon pair generation enabled by modal-phase-matched TFLN MRRs. Despite lower nonlinear efficiency compared to quasi-phase matching, modal phase matching may be more suitable for the integration of dense on-chip nonlinear quantum devices, such as state generation, manipulation, and detection. Advancements in fabrication technology, especially the development of ultra-high-Q TFLN MRRs, may enable the modal-phase-matched approach to achieve results comparable to periodically poled TFLN MRRs. This would thereby simplify fabrication, optimize performance, and facilitate easier integration into modern nonlinear and quantum photonic devices, expanding the potential applications of on-chip photon sources.

# Figures:

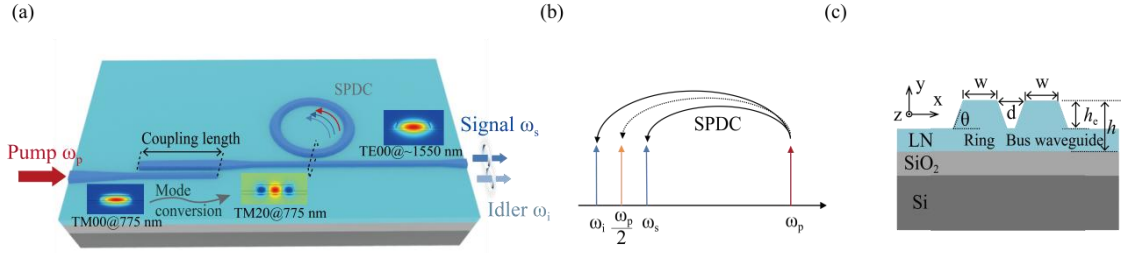

Fig. 1: On-chip TFLN MRR chip for photon pair generation. (a) Schematic diagram of the device, where the pump is sent into a straight waveguide, coupled and converted in mode converter, and then into the MRR, where SPDC takes place. Then the output signal and idler are coupled out and measured. (b) Schematic depiction of SPDC process, showing degenerate parametric down conversion (dashed line) and non-degenerate parametric down conversion (solid lines). $\omega_p$, $\omega_i$, $\omega_s$ are the pump, signal, and idler light of the SPDC process, respectively. (c) Schematic diagram of MRR waveguide cross-section (corresponding to the black dashed box in (a)). The TFLN layer thickness ($h$), the etching depth of the waveguide ($h_e$), and the sidewall angle ($\theta$) are 300 nm, 200nm, and 70°, respectively. d is the coupling gap between the bus waveguide and the micro-ring waveguide. The width of the bus waveguide ($w_{bus}$) is equal to the width of the micro-ring waveguide (w).

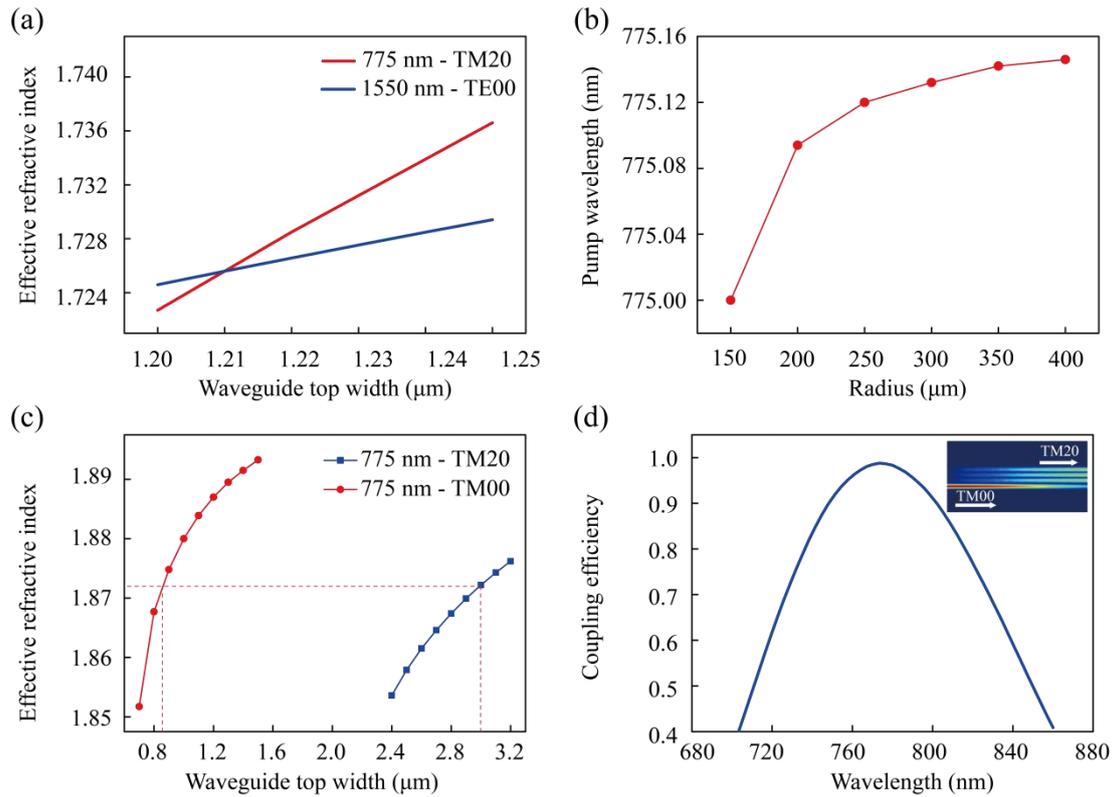

Fig. 2: Numerical calculations for modal phase-matching conditions and mode conversion. (a) The effective refractive index of 775 nm TM20 and 1550 nm TE00 as a function of waveguide top width at R = 150 μm. (b) Pump wavelength required for phase-matching as a function of different MRR radii for a waveguide top width of w = 1.21 μm. (c) Effective refractive index of two modes at 775 nm with varying waveguide top width of the mode converter. (d) The coupling efficiency versus input pump wavelength. The inset in (d) shows the modal field distribution of the mode converter at 775 nm.

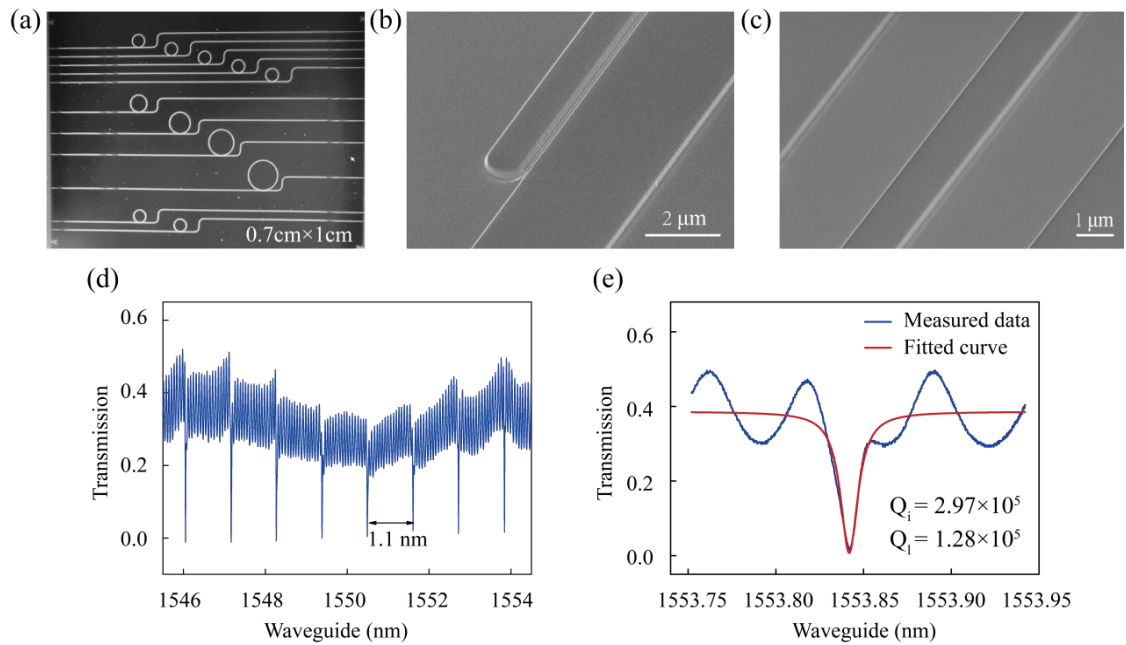

Fig. 3: Sample characterization. (a) Microscopic image of the MRRs. (b) SEM image of the mode converter. (c) SEM image of the waveguide. (d) Transmission spectrum of TFLN MRR for a radius of 150 μm and coupling gap of 0.45 μm, showing FSR of 1.1 nm for the MRR. (e) Transmittance spectrum around the resonant wavelength of the MRR in blue and the corresponding fitted curve in red indicating the intrinsic versus measured Q-factor.

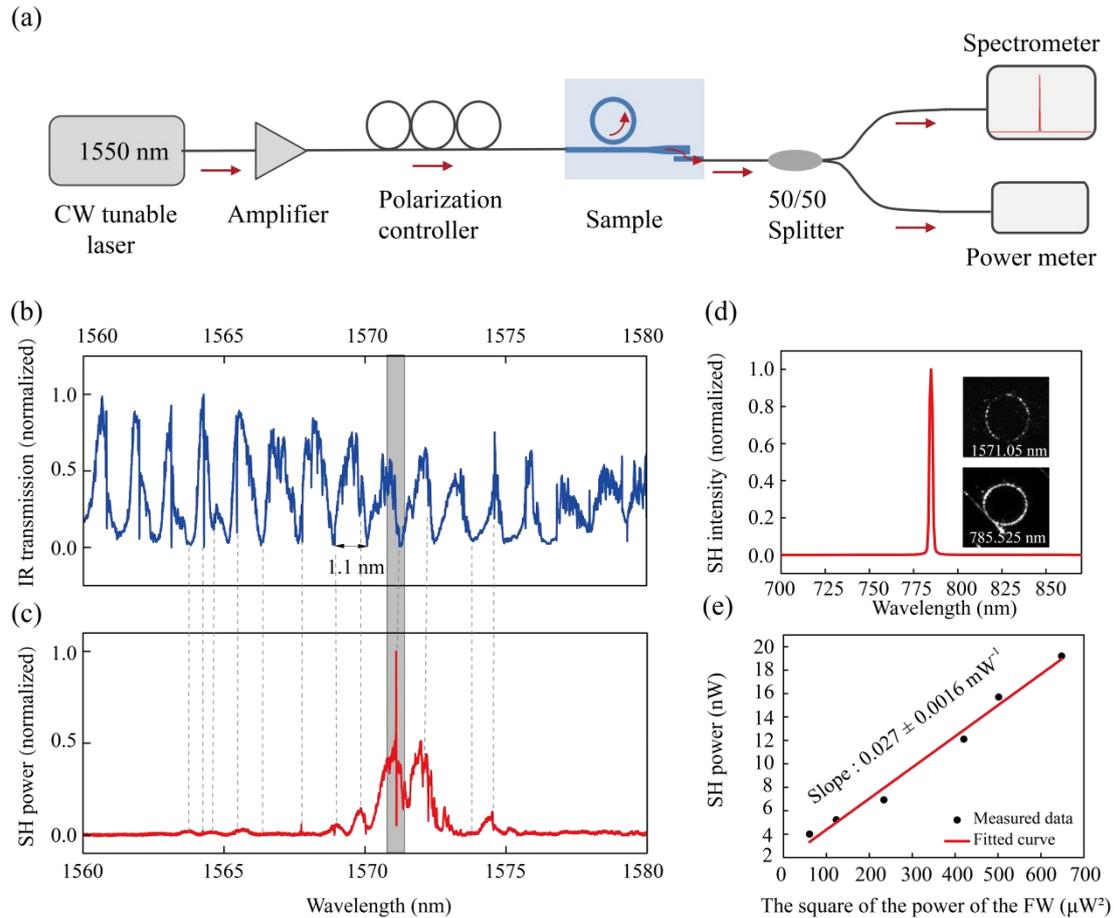

Fig. 4: Measurement results of SHG for the MRR (R = 150 μm, d = 0.45 μm) with a mode converter. (a) Schematic diagram of experimental setup for SHG. A tunable CW laser centered at 1550 nm is sent through an amplifier and polarization controller. The input fundamental light is coupled in the sample, and the output SHG light is sent through a 50/50 beam splitter to measure both the transmission spectrum, and the output power. (b) Measured transmission spectrum of the fundamental light signal through the sample, and (c) the corresponding SHG signal. (d) The SHG signal intensity for a fundamental wavelength of 1571.05 nm, corresponding to the fundamental and SHG signal's resonance and phase-matched conditions. The insets in (d) show the fundamental and SHG signal in the MRR captured by CCD cameras. (e) The SHG power as a function of the square of the fundamental light power, confirming the quadratic dependence (slope = $0.027 \pm 0.0016$ mW$^{-1}$).

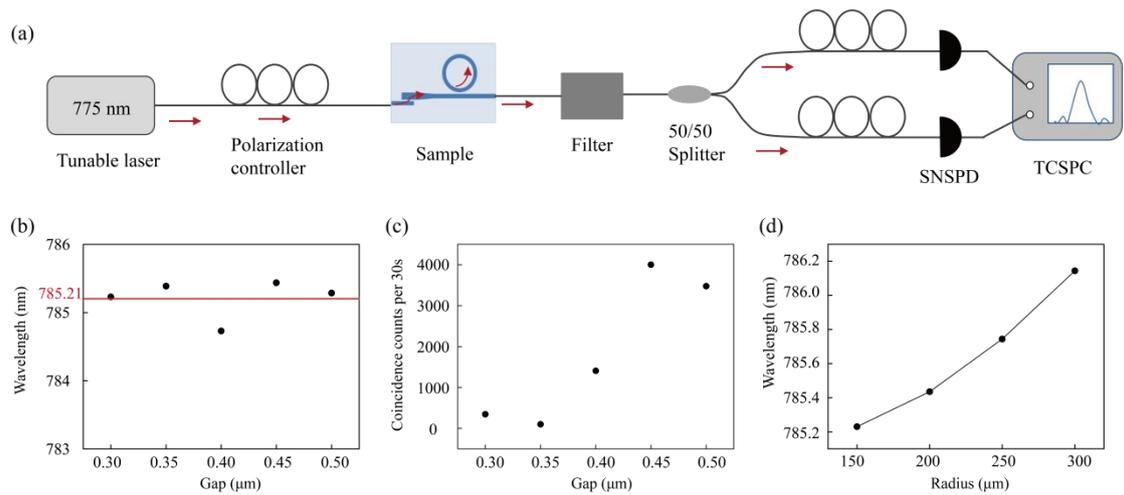

Fig. 5: Photon pair generation, detection, and measurement. (a) Schematic diagram of experimental setup for measuring photon pairs. A tunable pump laser at 775 nm is sent through a polarization controller and then coupled into the sample. The fundamental light is then filtered at the output such that only the generated signal is detected. The generated photon pairs are sent through a 50/50 beam splitter, subsequently sent through polarization controllers. Finally, single photon counts, and coincidence counts are measured using a time-dependent single-photon counter. (b) The pump wavelength for max coincidence count at different coupling gaps. The red line at λ=785.21 nm corresponds to near-degenerate photons generated by SPDC. (c) Coincidence counts for different coupling gaps for a 150 μm radius MRR. (d) Resonant pump wavelength for an MRR with a coupling gap of 0.3 μm for different radii.

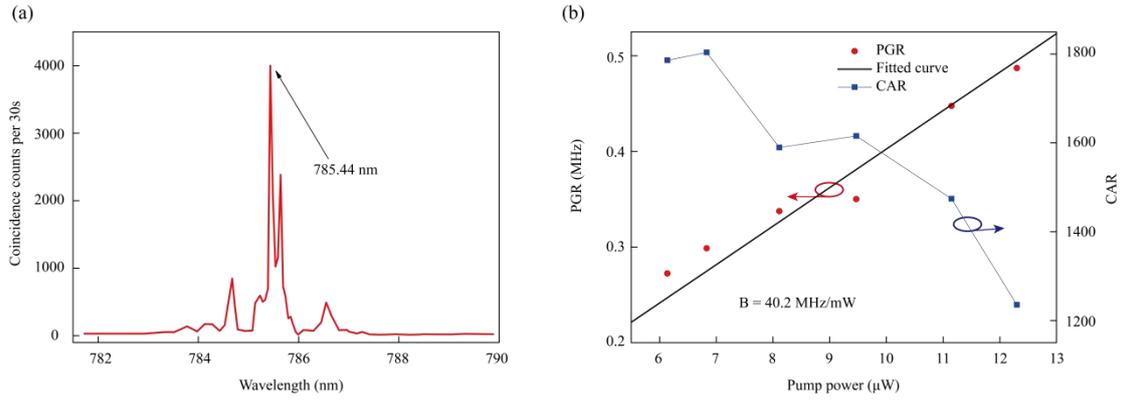

Fig. 6: Characterization of the quality of the generated photon pairs for a MRR of 150 µm radius and a 0.45 µm coupling gap. (a) Measured coincidence counts highlight the SPDC spectrum as a function of pump wavelength, indicating maximum coincidence for an input pump wavelength of 785.44 nm, consistent with the conditions for SHG. (b) Measured PGR in red, with corresponding fit curve in black, and CAR in blue for different input pump powers. The brightness, B, is found to be B = 40.2 MHz/mW.

Tab. 1: Performance comparison of photon-pair sources based on continuous-wave pumped micro-ring resonators with modal phase matching

| Platform | Radius (μm) | Type | Pump power | PGR (MHz) | Brightness @1 μW | CAR | Ref. |
|---|---|---|---|---|---|---|---|
| SOI | 73 | SFWM | 19 μW | 0.8 | 2.2 KHz | 600 | [40] |
| SOI | 10 | SFWM | 59 μW | 0.52 | 149 Hz | 532 | [41] |
| AlN | 30 | SPDC | 1.9 mW | 11 | 5.8 KHz | - | [42] |
| $Si_3N_4$ | 25 | SFWM | 46 μW | 10 | 4.8 KHz | 2200 | [43] |
| InP | 24 | SFWM | 22 μW | 0.07 | 145 Hz | 277 | [44] |
| AlGaAsOI | 13.91 | SFWM | ~20 μW | ~8 | 20 KHz | 353 | [45] |
| SiC | 43 | SFWM | 0.17 mW | 0.009 | 0.31 Hz | 600 | [11] |
| TFLN | 150 | SPDC | 12.3 μW | 0.49 | 40.2 KHz | 1235 | This work |